\documentclass[preprint2]{aastex}

\usepackage{threeparttable,textcomp,multirow}
\usepackage{tensor}
\usepackage{graphicx}
\newcommand{\simgt}{\,\hbox{\lower0.6ex\hbox{$\sim$}\llap{\raise0.6ex\hbox{$>$}}}\,}

\shorttitle{Testing the gravitational lensing explanation for the MgII
problem in GRBs}
\shortauthors{Rapoport et al.}

\begin{document}

\title{Testing Gravitational Lensing as the Source of Enhanced Strong MgII Absorption Towards Gamma-Ray Bursts}

\author{Sharon Rapoport\altaffilmark{1}, Christopher A. Onken\altaffilmark{1}, Brian
  P. Schmidt\altaffilmark{1}, J. Stuart B. Wyithe\altaffilmark{2},
  Brad E. Tucker\altaffilmark{1} and  Andrew J. Levan\altaffilmark{3}} 

\altaffiltext{1}{Research School of Astronomy and Astrophysics, The Australian
  National University, Weston Creek, ACT 2611, Australia}
\altaffiltext{2}{School of Physics, University of Melbourne,
  Parkville, Victoria 3010, Australia}
\altaffiltext{3}{Department of Physics, University of Warwick,
  Coventry, CV4 7AL, UK}
\begin{abstract}
  Sixty percent of gamma-ray bursts (GRBs) reveal strong MgII
  absorbing systems, which is a factor of $\sim$2 times the rate seen
  along lines-of-sight to quasars. Previous studies argue that the
  discrepancy in the strong MgII covering factor is most likely to be
  the result of either quasars being obscured due to dust, or the
  consequence of many GRBs being strongly gravitationally lensed. We
  analyze observations of quasars that show strong foreground MgII
  absorption. We find that GRB lines of sight pass closer to bright
  galaxies than would be expected for random lines of sight within the
  impact parameter expected for strong MgII absorption. While this
  cannot be explained by obscuration in the GRB sample, it is
  a natural consequence of gravitational lensing. Upon examining the
  particular configurations of galaxies near a sample of GRBs with
  strong MgII absorption, we find several intriguing lensing
  candidates. Our results suggest that lensing provides a viable
  contribution to the observed enhancement of strong MgII absorption
  along lines of sight to GRBs, and we outline the future observations
  required to test this hypothesis conclusively.
\end{abstract}


\section{Introduction} \label{intro} \cite{1538-4357-648-2-L93}
pointed out that gamma-ray bursts (GRBs) show approximately four times
as many MgII absorbing lines in their spectra as quasars, although the
current best estimate of the excess is now a factor of $\sim$2
(\citealt{2009AA.503.771V}; hereafter V09). Strong MgII absorbers
(equivalent width, EW, greater than 1\AA) are found towards $\sim$60\%
of GRBs with followed-up optical afterglows. Both being high redshift
beacons, GRBs and quasars might be expected to have similar lines of
sight through the cosmos, and explaining the preponderance of MgII
absorbers towards GRBs has proven a challenge \footnote{No difference
  between GRBs and quasars has been found for either weak MgII systems
  \citep{Tejos2009} or for CIV absorption systems \citep{Tejos2007}.}.

\cite{0004-637X-659-1-218} considered dust obscuration, beam size
differences, the intrinsic properties of GRBs, and gravitational
lensing as possible causes.  Beam size differences were found to be
irrelevant, with simulations predicting that the absorbing systems are
significantly larger than either the GRB afterglow ring or the quasar
accretion disk. If the additional strong MgII absorbers are physically
associated with the GRBs, one might expect the absorber properties to
be distinct from the systems towards quasars. However,
\cite{2009ApJ...697..345C} found no difference between the two
populations of absorbers. Ultimately, this leaves dust obscuration of
quasars and gravitational lensing of GRBs as the two most plausible
explanations.

If strong MgII absorbers are dusty, so that the discrepancy between
GRBs and quasars is due to quasars being preferentially lost from the
observed samples (GRB's being initially selected in gamma-rays are not
subject to this extinction), \citet{0004-637X-659-1-218} determined
that the number of MgII systems towards quasars would have to have
been underestimated by a factor of 1.3-2, which they found to be
unlikely (but see \citealt{2011arXiv1106.0692B}). Studies of quasars
have shown that the metal-enriched gas responsible for MgII absorption
is broadly associated with foreground galaxies
(e.g. \citealt{0004-637X-662-2-909}). Thus, if dust obscures quasars
with strong MgII, providing the origin of the GRB-quasar absorption
system discrepancy, then the $\sim$60\% of GRBs with strong MgII
absorbers should represent random lines of sight within 60\% of the
sky nearest to foreground galaxies. However, if lensing is responsible
for the discrepancy of MgII absorption between quasars and GRBs, then
there should be an excess of GRBs at small separations from foreground
galaxies. Therefore, the distribution of GRB-galaxy separations
provides a test to distinguish between the two hypotheses.

Finding multiple images of the same source is the calling card of
strong gravitational lensing. \citealt{2010arXiv1004.2081W} (W11)
explored the possibility that gravitational lensing combined with a
multi-band magnification bias (described in
\citealt{2003ApJ...583...58W}), could lead to the large number of high
equivalent width MgII absorbers for GRBs. W11 showed, if the gamma ray
and optical luminosities of GRBs are uncorrelated, and if the
luminosity functions have a cumulative slope with power law index $>
3.5$, then 10\%-60\% of the GRBs with afterglow follow-ups should have
been multiply-imaged. 

The V09 sample was chosen solely on the basis of the optical
afterglow's brightness and reveals 15 strong MgII absorbers out of the
26 GRBs sample. As W11 do not predict lensed GRBs to be
significantly brighter overall, we treat the V09 sample as an unbiased
group of GRBs and therefore expect 10\%-60\% of the 26 GRBs
should have been multiply-imaged. W11 also noted that, based on the sky
coverage of \emph{Swift}, the probability of the satellite detecting
three separate doubly-imaged GRBs is only 3\%. The probability that
none of the V09 sample was observed to be doubly-imaged, even if
lensing does occur, was found to be $\sim$50\%.

If GRBs are frequently strongly lensed, they open up a potentially
powerful probe of galaxies and the cosmos. Time delay and
magnification measurements, as done with quasar lenses, become
possible with exquisite precision, allowing accurate mass models and
distance measurements to be made to an ever increasing sample of
objects. 

In this paper, we investigate the possibility that
gravitational lensing is affecting a significant fraction of GRBs with
optical counterpart. For consistency with the W11 analysis, we study
archival {\it Hubble Space Telescope} (HST) observations of the galaxies near the lines of sight
towards each of the GRBs in the V09 sample, and estimate the
probability that the GRB was strongly lensed. In section \ref{data}, we describe the data used and our data
reduction procedures. In section \ref{Galaxy Proximity to GRB
  Lines-of-Sight}, we study the probability of finding nearby galaxies
towards GRBs with strong MgII absorbers vs.\ a random line of
sight. In section \ref{Lensing Analysis of Individual GRBs}, we
outline the lensing analysis methods and properties of the model employed, and
present our results for the individual GRBs. We discuss our
conclusions in section \ref{discussion}. Throughout this paper we
assume a cosmology of $\Omega_M=0.27$, $\Omega_\Lambda=0.73$ and
$H_o$=72 km/s/Mpc.

\section{Data} \label{data}

In order to consistently test the W11 hypothesis, we analyze the same
26-GRB sample from V09, from which their model is based. As we are
searching for galaxies with small impact parameters, we require high
spatial resolution, which is best obtained with HST. In the few cases
where HST data are not available, we use data from 8-m class
telescopes. Due to the relatively poor angular resolution of the VLT
and Gemini imaging data, a null detection in these cases does not rule out
lensing. These data are taken as a part of our statistics only when a
possible lensing galaxy is observed.

Calibrated HST images available from the archive, taken with the STIS
and ACS instruments, were combined in python by way of the
STSDAS/\emph{Multidrizzle}
package\footnote{http://www.stsci.edu/hst/HSToverview/documents/multidrizzle}. VLT
ISAAC, FORS1 and FORS2 images were similarly reduced using the relevant
pipeline within the
\emph{gasgano}\footnote{http://www.eso.org/sci/software/gasgano/}
software package. Version 1.9 of the Gemini IRAF
package\footnote{http://www.gemini.edu/sciops/}
was employed in reducing data from Gemini's GMOS instrument
\citep{2004PASP.116.425H}. After stacking all images from each filter
and constructing deep images from our reduced data, nearby galaxies
($< 5^{\prime\prime}$) were located using SExtractor
\citep{1996AAS.117.393B}. The impact parameter between these galaxies
and their partner GRBs was subsequently measured.

In total, we have deep images of 11 objects with strong MgII absorbers
from the sample of V09 (representing $\sim$75\% of the 15 GRBs with
strong MgII absorbers). Since the HST observations were not made on
the basis of strong MgII absorption, we believe that this is an
unbiased sample for this analysis.

\begin{deluxetable}{llllccccp{2cm}}
\tabletypesize{\footnotesize}
\tablewidth{20.5cm}
\rotate 
\tablecaption{GRB Field Imaging Results\label{tab1}\tablenotemark{1}}
 \tablehead{
 \colhead{GRB} &  \colhead{Instrument} & \colhead{Program
   ID}&\colhead{Filter} & \colhead{Limiting
   Magnitude\tablenotemark{2}} & \colhead{Galaxy Separations (\tablenotemark{3})} &
 \colhead{$m(AB)$}&\colhead{GRB redshift}&
\colhead{Strong MgII \tablenotemark{4}}\\
 \colhead{} & \colhead{} & \colhead{} &
 \colhead{} &\colhead{(AB mag)} &\colhead{(arcsec)}& \colhead{}& \colhead{} &\colhead{redshift}
           }
 \startdata
 020405 & HST/WFPC2 & HST 9180      & F702W   & 23.8
 & 2.6 (1) & 21.7 $\pm$ 0.3 &0.695 & 0.472\\
 030429 & VLT/FORS2 & VLT 71.D-0355 & R\_BESS & 22.0\tablenotemark{5} & 1.2 & 22.7 $\pm$ 0.1\tablenotemark{6}&2.66 &0.814\\
 021004 & HST/ACS   & HST 9405      & F606W   & 26.1                  & 1.4 (4)& 26.8 $\pm$ 0.4 &2.33& 1.3800, 1.6026 \\
 \nodata & \nodata   & \nodata       & \nodata & \nodata              & 1.8 (3)& 26.8 $\pm$ 0.4 & \nodata & \nodata \\
 \nodata& \nodata         & \nodata   & \nodata   & \nodata           & 3.2 (5)& 26.5 $\pm$ 0.4& \nodata& \nodata  \\
 \nodata& \nodata         & \nodata   & \nodata   & \nodata           & 3.7 (2)& 26.1 $\pm$ 0.3& \nodata & \nodata \\
 \nodata& \nodata         & \nodata   & \nodata   & \nodata           & 4.3 (1)& 25.9 $\pm$ 0.3& \nodata & \nodata \\
 010222 & HST/WFPC2 & HST 8867& F606W         & 22.4                  & 3.6 (3)& 25.7 $\pm$ 1.0 &1.477 &0.927, 1.156 \\
 \nodata & \nodata   & \nodata       & \nodata & \nodata              & 3.9 (4)& 25.0 $\pm$ 0.7 & \nodata & \nodata \\
 \nodata & \nodata   & \nodata       & \nodata & \nodata              & 4.0 (11)& 23.1 $\pm$ 0.3 & \nodata& \nodata \\
 060206  & HST/WFC  & HST 10817  & F814W  & 23.9
 & 0.9 (1) & 26.4 $\pm$ 0.5 &4.048&2.26\\
 \nodata & \nodata   & \nodata       & \nodata & \nodata
 & 2.4 (2) & 24.8 $\pm$ 0.2 & \nodata& \nodata \\
060418 &  HST/ACS  & HST 10551 & F775W   & 25.5 & 1.1 (3)&24.2 $\pm$ 0.2&1.49     & 0.6026, 0.6559, 1.1070\\
 \nodata & \nodata   & \nodata       & \nodata & \nodata & 1.5 (2)& 25.2 $\pm$ 0.3 & \nodata& \nodata \\
 \nodata & \nodata   & \nodata       & \nodata & \nodata & 3.5 (1)& 22.5 $\pm$ 0.1 & \nodata& \nodata \\
 050820A&  HST/ACS& HST 10551& F625W   & 25.4 &3.3 &24.7 $\pm$ 0.2&2.615& 0.9615, 1.4288\\
 \nodata & \nodata   & \nodata       & \nodata & \nodata
 & 3.5 & 26.5 $\pm$ 0.5 & \nodata& \nodata \\
 080319B& HST/WFPC2 & HST 11513& F606W   & 24.4 &1.7&25.1 $\pm$ 0.7&0.9378&0.7154 \\
 \nodata & \nodata   & \nodata       & \nodata & \nodata              & 2.6 & 24.9 $\pm$ 0.6 & \nodata & \nodata \\
 \nodata & \nodata   & \nodata       & \nodata & \nodata              & 4.0 & 25.0 $\pm$ 0.7 & \nodata& \nodata \\
 991216 & HST/STIS  & HST 8189 & Clear   & 26.1&0.4&24.5 $\pm$ 0.3 &1.022&0.770, 0.803\\
 \nodata & \nodata   & \nodata       & \nodata & \nodata & 2.4 & 23.7 $\pm$ 0.2 & \nodata& \nodata \\
 020813& HST/ACS  & HST 9405 & F606W   & 26.7 & 2.3& 24.1 $\pm$ 0.1&1.255&1.224\\
 \nodata & \nodata   & \nodata       & \nodata & \nodata & 4.0 & 24.1 $\pm$ 0.2 & \nodata & \nodata \\
 050908 & HST/ACS  & HST 11734& F775W & 26.1 &1.0& 26.6 $\pm$ 0.6&3.55&1.548\\
 \nodata & \nodata   & \nodata       & \nodata & \nodata & 2.8 & 25.7
 $\pm$ 0.4 & \nodata& \nodata \\
 \nodata & \nodata   & \nodata       & \nodata & \nodata & 3.7 & 25.7 $\pm$ 0.4 & \nodata & \nodata \\
 \enddata
 \tablenotetext{1}{Table includes only nearby ($<5^{\prime\prime}$) galaxies which are not
   confirmed to be a host}
 \tablenotetext{2}{10-sigma within an aperture of 0.2 square-arcseconds,
   observed limit for each combined image.}
 \tablenotetext{3}{Galaxy ID number when an image is provided within
   this paper}
 \tablenotetext{4}{Taken from V09}
\tablenotetext{5}{10-sigma for a point source}
\tablenotetext{6}{Jakobsson et al. 2004}
\end{deluxetable}

\section{Galaxy Proximity to GRB Lines-of-Sight} \label{Galaxy Proximity to GRB Lines-of-Sight}

\cite{0004-637X-691-1-152} noted the consistent presence of galaxies
at small angular separations from a sample of four GRBs showing
strong MgII absorbers, and the absence of such galaxies near three GRBs
without them. Such a correspondence is not surprising since the
MgII absorption is thought to arise from material associated with
galaxies. Our analysis takes this approach further by roughly
tripling the sample size, and by comparing MgII lines of sight to the
more robust baseline of a distribution constructed from random lines
of sight.

According to V09, $\sim$60\% of all GRBs have
a strong MgII absorbing system ($>1$\AA). Using the galaxy catalog of
\cite{2006AJ.132.926C} for the Hubble Ultra Deep Field (UDF;
\citealt{2006AJ.132.1729B}), we measure the radius around every galaxy
brighter than a given magnitude which, together, would give 60\% sky
coverage using a top hat function; we denote this radius as
$\theta_{MgII}$. This top-hat model represents the most extreme
concentration possible for MgII absorption along random lines of
sight, and thus is the most stringent comparison for the observed
impact parameter distribution. It may not, however, be an unrealistic
model, as \cite{2010ApJ.714.1521C} used quasar sight-lines to measure a
high MgII covering fraction within a certain (luminosity-dependent)
galaxy radius, and found that the covering fraction fell sharply at
larger radii.

If GRBs are randomly distributed, one would expect the 60\% of GRBs
which have MgII absorption to have a random distribution of
separations from galaxies within the radius $\theta_{MgII}$. However,
if the cause for the discrepancy between GRBs and quasars is due to
gravitational lensing, we would expect the distribution of galaxies
around GRBs with strong MgII lines to be more concentrated
towards small radii. In the framework of our UDF analysis, the value
of $\theta_{MgII}$ in any filter is a function of only the limiting
magnitude used to constrain the galaxy catalog. 

\subsection{Distribution of separations to nearby galaxies}
We use a Monte Carlo approach to generate random lines of sight
through the UDF, and measure the distances to the nearby galaxies in
the Coe et al.\ catalog. We adopt a limiting magnitude for the catalog
galaxies of 27.5 mag (AB) in each filter, since this corresponds to
the faintest nearby galaxy in the GRB images. Running 10,000 realizations in
each of the four UDF filters, we then weight the relative
contributions from each filter to match the filter distribution for
the GRB images (using only the filter that gives the deepest stacked
image for each GRB in the analysis, and taking the UDF filter most
similar to that used with the GRB).

Our statistical analysis does not distinguish between galaxies that
lie in front of the GRBs and those behind. As long as the galaxy
number counts continue to rise at the faint end (a condition satisfied
for F775W=27.5 mag [Fig.~29 of Coe et al. 2006], as well as for the
other UDF filters), the consequence of our blindness to the galaxy
redshifts is to add spurious objects to our analysis for both the GRBs
and the random lines of sight. This will weaken the signature of
lensing by galaxies at small impact parameters by mixing in unrelated
background objects and reducing the difference in the cumulative
radial distributions. If MgII absorption is related to galaxies that
are typically fainter than any of our images, then we would expect the
spatial distribution of galaxies to be the same for the GRBs and the
random lines of sight.  \cite{2004ApJ...614...84B} and
\cite{2009MNRAS.400.1613C} found that long-GRB host galaxies populate lower density regions than
average, and we therefore do not expect a bias due to galaxies
at the redshift of the GRB.

A Kolmogorov-Smirnov (K-S) two-sample test rejects the hypothesis of
the GRB data being drawn from the random distribution, but only at
$>90\%$ confidence when using a limiting magnitude of 27.5; because we
would not expect the entire population to be lensed, but rather a few
very unusual cases, it is not surprising that this treatment is
inconclusive.

\subsection{Probabilities of nearest bright galaxy}

Assuming that lensing is caused by the nearest galaxy, we perform a
test to compare the probability of the 11 different alignments between
the GRB and its \emph{nearest} galaxy. This time, we find the
probability of each event happening when comparing to a line of sight
near a galaxy with the same apparent brightness as our GRB
host. While, as mentioned above, GRB host galaxies are not likely to
reside within a cluster, this
is done to account for any improbable clustering, and falsely
identifying galaxies that are associated with the GRB redshift itself.

\begin{figure}[htb]
\begin{center}
\includegraphics[width=\columnwidth]{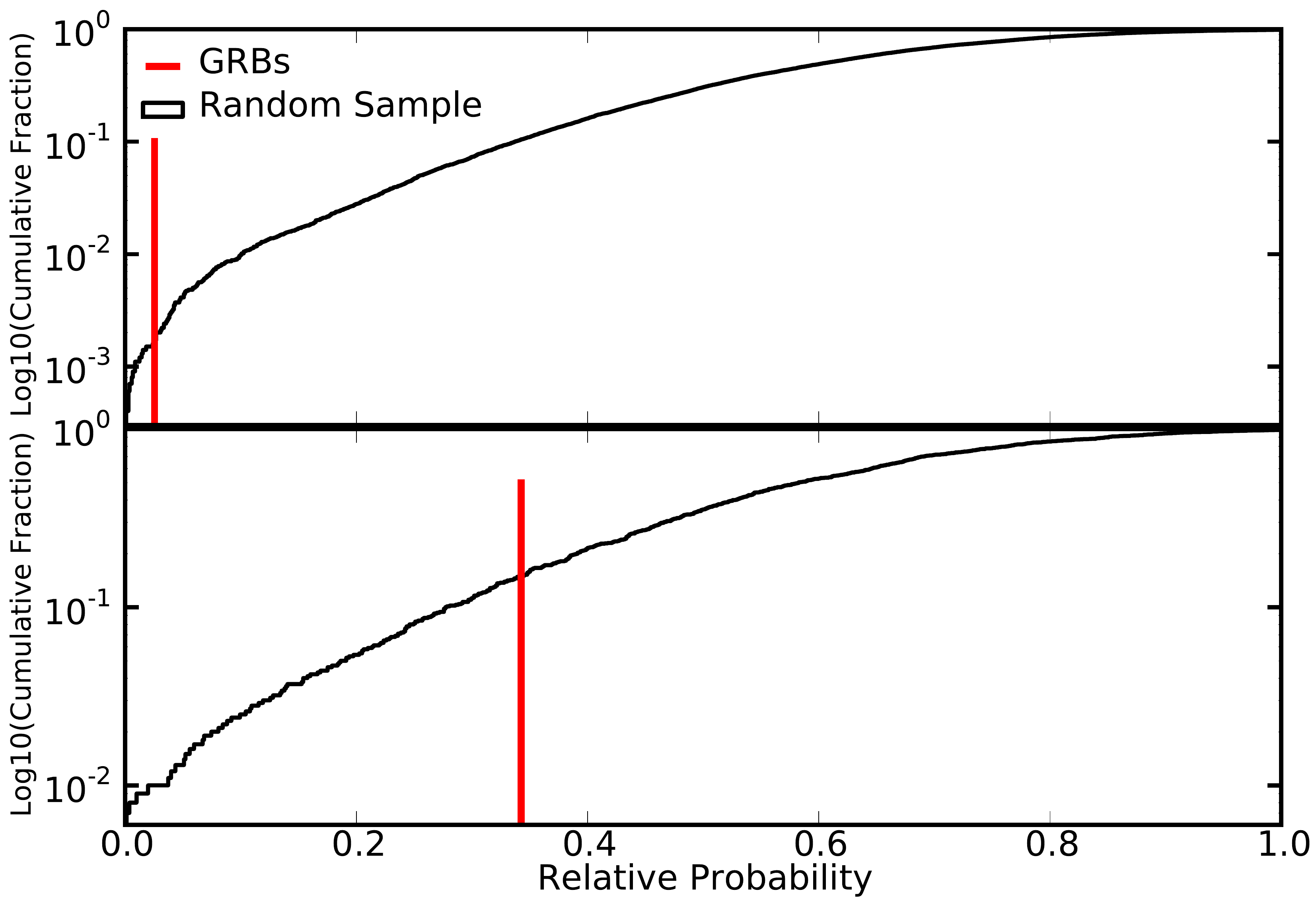}
\caption[short]{Monte-Carlo analysis describing the probability of
  having the alignments between the GRBs and their \emph{nearest}
  galaxy. The cumulative fraction is found using a Monte-Carlo
  approach, and choosing the nearest 11 galaxies towards random
  galaxies having the magnitudes of the GRB hosts ($\pm$0.5). Where a
  GRB host is not observed, a random line of sight is chosen. The
  upper panel is for the 11 GRBs, showing a probability of 0.2\%. The
  lower panel shows the result after removing the 3 most
  likely candidates: GRB020405, GRB030429 and GRB991216. The
  probability of having the alignment of the 8 GRBs left is 15\%.}
\label{fig:prob}
\end{center}
\end{figure}

After randomly choosing a galaxy from the UDF with similar apparent magnitude
as the actual GRB host ($\pm$0.5 mags), we measure the magnitude of and
distance from it to the nearest galaxy (up to a specific limiting
magnitude). For GRBs without known host galaxies we use a random line
of sight. The probability of finding a closer and/or nearer galaxy is
calculated and the probabilities of the 11 alignments are multiplied
together to form one probability for the ensamble. It is important to
note that only alignments with a galaxy within $\theta_{MgII}$ are
counted (as we assume those galaxies are the cause for the strong MgII absorption). In
figure \ref{fig:prob}, the histogram of the multiplied probabilities
is plotted (for limiting magnitude of 27.5 in V band). The red line
represents the probability of the GRBs alignments. We find the
probability of having a set of alignments as close and as 
bright as the observed sample of 11 GRBs is only 0.2\% (upper
panel). To verify that random lines of sight are representative of the
GRB population without strong MgII absorption, we compare the random
lines of sight result to that of 9 GRBs without MgII absorption which
have publicly available HST observations (060526, 061201, 070721B,
060313, 050730, 050419A, 050525, 060614 and 060729). The probability of finding
the alignments from these 9 GRBs to their nearest galaxy is 78\%,
confirming that our MC routine is adequate. 

As we remove our probable candidates for strong lensing and possibly magnified GRBs from the
analysis (see \S \ref{Lensing Analysis of Individual GRBs}), we expect
the alignments to be similar to our expectation from a random line of
sight. Therefore, we test the outcome when ignoring the most likely
candidates GRB020405, GRB030429 and GRB991216. The same simulation
now finds the alignment probability of the 8 remaining GRBs to be
15\% (lower panel). As the distribution appears to have a
log-normal shape, P=0.15 is just within 1 standard deviation of the
mean (P=0.5) for a random line of sight. Following the statistical
approach of V09, we find that the absorber density per unit redshift ($\frac{\partial
  n}{\partial z}$) changes from $0.74\pm0.20$ to $0.58\pm0.20$ when
removing GRBs 020405, 030429 and 010222. When
comparing to the QSO result of $0.278\pm0.010$ \citep{Nestor2005}, the significance of the
MgII discrepancy is reduced by one sigma. However, GRBs are not found to
completely agree with the absorbers distributions in QSO, implying
another process (e.g. dust obscuration) must be involved.

In order to test the effect of the limiting magnitude, we repeat the
analysis for limiting magnitudes in the range 27.0 - 28.5. The
brighter limit is chosen to match the faintest nearby galaxy observed
for the GRBs, and the fainter is chosen as a highly conservative
limiting magnitude for our images (we do not have such deep images,
however a fainter limiting magnitude will make $\theta_{mgII}$ smaller,
which makes it less likely to find a galaxy as bright or as close as
the nearest galaxies to the GRBs). The results for the probability of the 11 GRBs happening
range from 0.1-0.4\% (and for ignoring the likely cases range from 12-16\%),
suggesting that our analysis is not highly sensitive to the limiting
magnitude. This confirms that GRBs have brighter, closer galaxies than
expected if the 60\% of GRBs with MgII represent random lines of sight
within the 60\% of the sky closest to foreground galaxies. This
is as expected under the gravitational lensing
hypothesis.

To verify our results are not highly dependent on cosmic variance, we
perform another test. Taking the extreme case (galaxies with $\sim
10^{11}M_\odot$) for the UDF from \cite{cosmic}, we assume 30\%
variance 
within the whole UDF field. In a Monte-Carlo simulation we increase
the number of galaxies in the field drawing from a gaussian with a
mean taken from the size of the UDF catalog and 30\%
variance. Each extra galaxy is randomly placed in the field and is
given a magnitude which is randomly drawn from the UDF magnitude
distribution. Repeating the process above we find the probability of
the 11 GRBs having their alignments with their nearest galaxy to
increase to only 3\%. We therefore conclude that our results are not
particularly sensitive to cosmic variance.

\section{Gravitational Lensing Models of individual GRBs} 
\label{Lensing Analysis of Individual GRBs} 

We showed above that GRBs with strong MgII absorbers include outliers
with improbable small separations from foreground galaxies. We
therefore conclude that lensing is a likely cause of the MgII absorber
excess since it is consistent with such an alignment, whereas
obscuration of quasars is not. Armed with this motivation, we
investigate the individual cases to look for potential lenses. In this
section we proceed under the assumption that the galaxy proximity is
related to gravitational lensing of the GRBs and conduct detailed
studies of the lensing likelihood for each GRB in the V09 sample.

It is important to note that like our statistical analysis, the
examination of lensing for individual GRBs may be confused by the
presence of galaxies lying behind the GRB in question. While we can
use the UDF redshift catalog of Coe et al. (2006) to constrain the
likely distances to some of our galaxies based on their SEDs,
determination of the actual redshifts can only be accomplished by
future observations.

We determine the properties of all potential lensing galaxies using
the methods of Tucker et al. (2012a,b in prep). In summary, galaxy
spectral energy distribution (SED) templates are created from the UV
to Infrared using 11 galaxies that cover a complete range of
morphological types \citep{CKSB94,Calzetti96,Kinney96,IRTemplates}. We
reproduce the range of observed galaxies via linear combinations of
all SEDs, and find all acceptable fits to the observed photometry
using a $\chi^2$ statistic. From the range of acceptable SEDs, we
calculate rest-frame magnitudes and the $\chi^2$-weighted galaxy
classification when data in several bands is available.

In the few cases where we identified the galaxy as a spiral and could
estimate M$_B$, we use the Tully-Fisher relation
\citep{2011arXiv1102.3911M} to approximate the rotational velocity of
stars in the lensing galaxy. In order to convert from the rotational
velocity to velocity dispersion, we divide by $\sqrt{2}$
\citep{binney1988galactic}. Incorporating the angular impact
parameter, the estimated velocity dispersion, and the known absorber
redshifts into a singular isothermal sphere (SIS) model, we use the
observational and model uncertainties to calculate the probability of
magnification and of the lens producing two images of the GRB.

We use the GRAVLENS/LENSMODEL software \citep{2001astro.ph..2340K} to
model the complex systems (i.e., those with multiple lensing
galaxies). The software allows one to specify the locations of the
source (GRB), lenses, and images of the source (including time delay,
when applicable), to constrain possible mass profiles of the
lenses. The output includes the image magnifications and time delays
(when not provided), and the critical curve of the lenses in the image
plane. In the few cases where more than one galaxy is observed near
the GRB and where none of the galaxies is obviously associated with
the absorbing system (via a spectroscopic redshift for the galaxy or
extreme close proximity to the GRB), we model them as a group where
each galaxy is associated with a Singular Isothermal Ellipse (SIE)
with shear. Where there is only one nearby galaxy, and no other
observational constrains, such as a second image, we assume a Singular
Isothermal Sphere (SIS) mass profile for the lensing galaxy. This
model predicts multiple images where the impact parameter, $\theta_i$,
is smaller than twice a galaxy's Einstein radius, $\theta_E$, which is
defined as
\begin{equation}
\label{eq.EinsteinRadius}
\theta_E = 4\pi\left(\frac{\sigma_{\nu}}{c}\right)^2\frac{D_{ls}}{D_s}.
\end{equation}
Here $\sigma_{\nu}$ is the line-of-sight velocity dispersion of the
galaxy (which could be constrained from the TF analysis described
above), and $D_{ls}$ and $D_s$ are the angular diameter distances
between the lens and the source, and the observer and the source,
respectively \citep{Schneider}. In addition to lensing geometry, the
model also provides information regarding intensity magnification and
time delay between images. In the simplest case of a single lensing
galaxy, the first image is further from the center-of-mass of the lens
and is brighter than the second image, making it more likely to be
detected. More complex lensing systems are not bound by such
constraints, but still make predictions about magnifications and time
delays.

Although MgII absorbers and intervening galaxies are correlated for
quasar sight-lines, the relation between the impact parameter and the
MgII equivalent width is not a tight one \citep{2005pgqa.conf.24C},
making the task of identifying the right absorber for each system not
straight-forward. As we normally lack the data to verify the nearby
galaxies' redshifts, we implement a statistical test to understand the
likelihood of a nearby galaxy being at the absorber's redshift. Again
using the UDF galaxy catalog of \cite{2006AJ.132.926C}, we perform a
Monte-Carlo simulation that measures the probability of having a
galaxy with a given apparent magnitude within a known distance from a
random line of sight. This allows us to estimate the chances of
finding galaxies near a GRB.

Hereafter, we analyze the possibility of strong gravitational lensing
for each of our GRBs, in descending order of multiple-lensing
likelihood.

\subsection{Candidates of Multiply Imaged GRBs}
\subsubsection{\emph{GRB020405}}
\begin{figure}[htb]
\begin{center}
\includegraphics[width=\columnwidth]{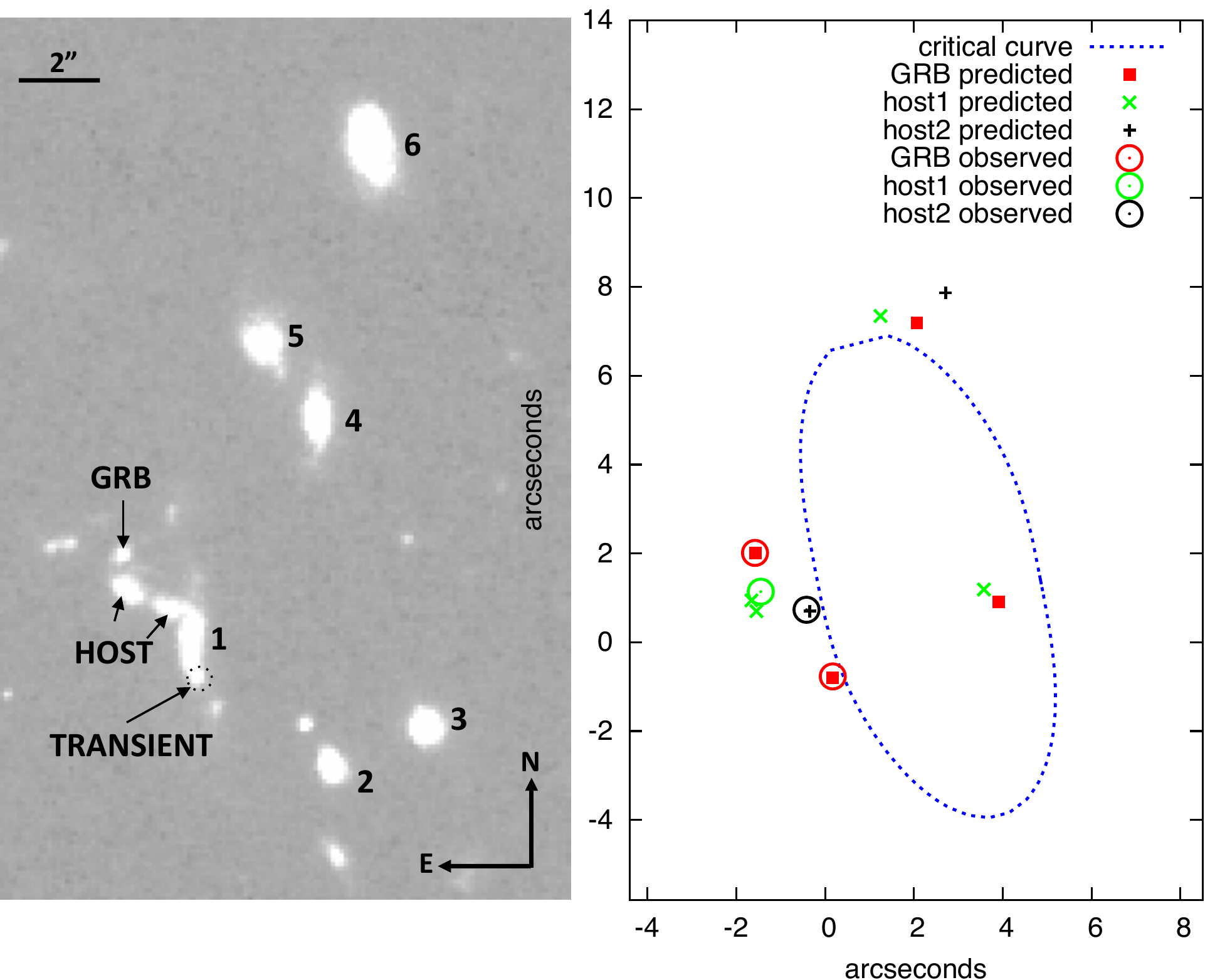}
\caption[short]{\emph{Left:} HST WFPC2/F702W field of GRB020405. The
  GRB is clearly visible and the complex host and second transient are
  indicated. Objects 1 and 2 are confirmed to be at the strong
  absorber's redshift of z=0.472. Galaxies 3-6 may be part of a group at the
  absorber's redshift. \emph{Right:} A representative LENSMODEL solution for SIE with
  shear. The time delay between the GRB images is $\sim$120 days where
  the leading image is $\sim$1.8 times brighter that the later
  one. The two predicted GRB images on the right arrive hundreds of days
  earlier. The host images to the right are less magnified.}
\label{fig:grb020405model}
\end{center}
\end{figure}

HST images of this GRB (z=0.695) revealed another transient object $3
^{\prime\prime}$ away, as noted by \cite{grb020405}. A VLT spectrum
confirmed that both the nearest galaxy and one other (objects 1 and 2,
respectively, in Figure \ref{fig:grb020405model}) are at the redshift
of the z=0.472 MgII absorbing system. The second transient
is visible along the edge of galaxy 1. Objects 3,4,5 and 6 are nearby
galaxies which we speculate below could possibly be part of a group at
the absorber redshift.

\begin{figure}[htb]
\begin{center}
\includegraphics[width=\columnwidth]{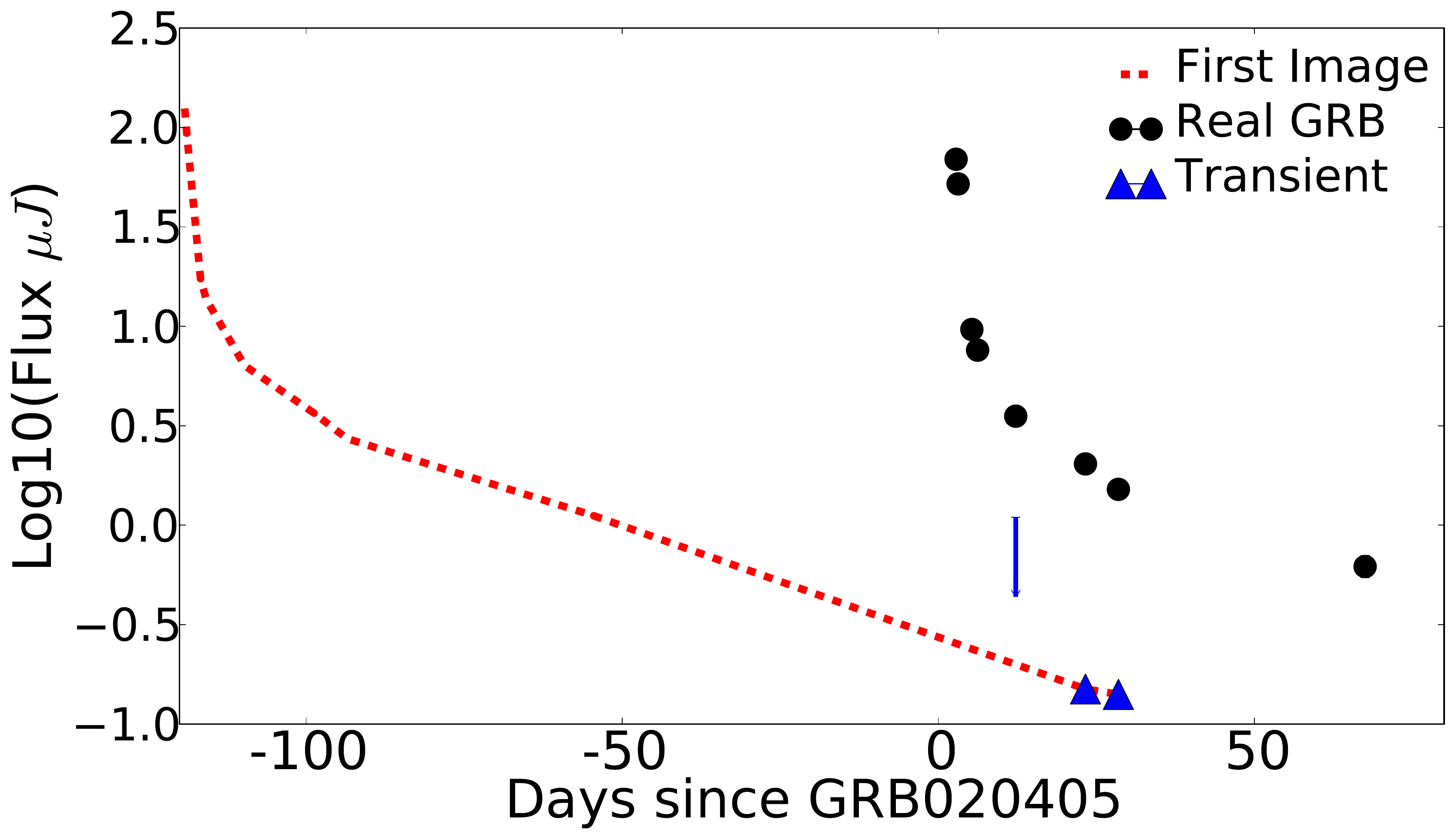}
\caption[short]{Light curve prediction for the lensing model. In black
dots, the observed I band flux of GRB020405. In red, the predicted
light curve of the first image, magnified by 1.8 with a time delay of
-120 days. The blue dots are the observed flux of the second transient
in the HST images, and the blue upper limit is found when subtracting
the last HST image from the early VLT I image.}
\label{fig:grb020405curve}
\end{center}
\end{figure}

The second transient was observed in the first observation of HST, 23
days after the GRB trigger, and was fainter than the GRB at all
times. With only few data points for this transient, it is consistent
(within errors) with the expected colour evolution of the GRB. While
we cannot dismiss the possibility of the second transient being a
non-related field SN, we expect the probability of finding an object
3$\prime\prime$ away from any $z\sim0.5$ redshift galaxy is low.
Modelling this system with LENSMODEL, we are able to explain the
unknown transient as an earlier, brighter image of the GRB (Rapoport
et al. in prep). The model assumes galaxies 1-6 are a group at the
MgII absorption redshift. We assume the galaxies lie in common dark
matter envelop and use a Singular Isothermal Ellipse (SIE) mass
profile with shear. While the time delays, positions of images and
their magnification are highly sensitive to the model, possible
solutions include a time difference between the images of $\sim$120
days, with the first GRB image being 1.8 times brighter. One such
model is shown in figure \ref{fig:grb020405model}. The transient
(i.e., first GRB image) would not have been detected in optical at the
time of the original GRB020405 observations, as the resolution and
depth required to separate the transient from the nearby galaxy were
not achieved by the ground-based facilities used in the first few
weeks after the gamma-ray trigger. Figure \ref{fig:grb020405curve}
shows the GRB observations in I band in black and the predicted flux
for the first image in red (1.8 times brighter). The blue points are
the observed fluxes of the second transient from the HST images and
the upper limit was found by subtracting the last HST observation from
the VLT I band image. (The earlier epochs for GRB020405 were obtained
with smaller telescopes, and provide no additional constraints on the
light curve of the second transient.). We are pursuing further
observations to verify the redshift of the galaxies in the field. The
model predicts several less-magnified images of the host complex,
which do not conflict with current observations. In addition, some
models suggest there were even earlier images of the GRB (on the right
of Fig. \ref{fig:grb020405model}), which occurred years before. It is
not possible to explain the second transient as a GRB image if the
nearby galaxies are not at the absorbers redshift. Phot-Z analysis,
using the EASY software
\citep{2008ApJ.686.1503B}, of
the possible group members shows consistency for the galaxy redshifts,
but with large uncertainty.

\subsubsection{\emph{GRB030429}}
\begin{figure}[htbp!]
\begin{center}
\includegraphics[width=\columnwidth]{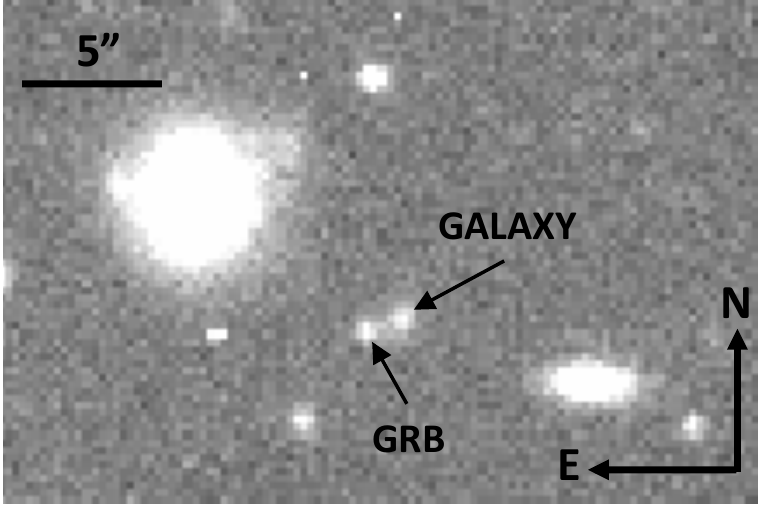}
\caption[short]{VLT/FORS2 R\_BESS field of GRB030429, with
  the potential lensing galaxy $\sim$1.2$^{\prime\prime}$
  to the right.}
\label{fig:grb030429}
\end{center}
\end{figure}

The impact parameter between the GRB030429 and the nearby galaxy is
1.2$^{\prime\prime}$ (figure \ref{fig:grb030429}), and the redshift of
the galaxy was confirmed by \cite{2004AA.427.785J} to be that of the
absorbing MgII system. Moreover, at a redshift of 2.66 with an
absorbing system at a redshift of 0.8418, the angular diameter
distance ratio $\frac{D_{\rm l}}{D_{\rm ls}}$ is approximately
0.54. As gravitational lensing is most efficient when the angular size
distance of the source as seen by the lens equals the angular size
distance between the lens and the observer, GRB030429 represents a
likely candidate for being lensed.

\begin{figure}[htbp!]
\begin{center}
\includegraphics[width=\columnwidth]{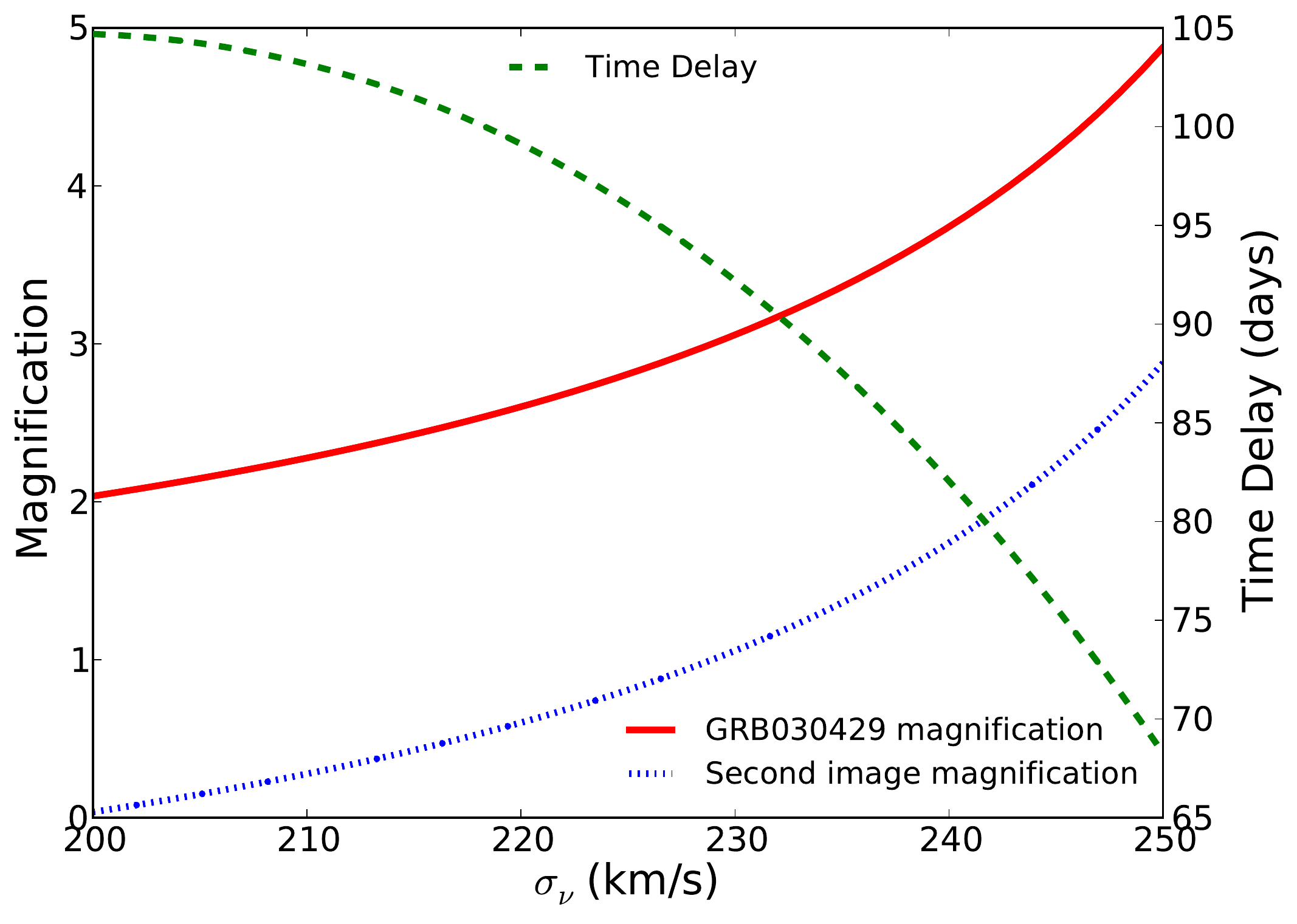}
\caption[short]{The prediction from a SIS model for the magnification
  of GRB030429 (blue), its second image (green), and the time delay
  between them (red).}
\label{fig:grb030429plot}
\end{center}
\end{figure}

Our Monte-Carlo simulation with the UDF finds the probability of
having a galaxy with the observed magnitude within
$1.2^{\prime\prime}$ along a random line of sight to be 0.4\%. For the
lines of sight towards our 11 GRBs, the chances of randomly finding
one such alignment is $\sim$14\%. Using the magnitudes of the
galaxy given by \cite{2004AA.427.785J}, our SED fitting techniques
imply $M_B = -21.1 \pm 0.1$. The TF relation for such an intrinsically
bright galaxy suggests a velocity dispersion of $160 \pm 65$ km/s
which corresponds to $\theta_E = \tensor*{0.38}{^{+0.37}_{-0.25}}$
arcseconds. With a separation of 1.2 arcsec, the SIS model predicts a second image for this GRB only if
$\sigma_{\nu} \simgt 200$ km/s. Since lensing would select out those
galaxies with the highest $\sigma_{\nu}$, the TF relationship is
useful to show plausibility in this case. Because the Einstein radius
for this system needs to be $\simgt 0.6^{\prime\prime}$ for multiple imaging, the
impact parameter of the observed GRB means it was likely the first of
two images if lensed. Figure \ref{fig:grb030429plot} shows the
expected magnifications of the GRB and the
predicted second image, and the time delay between the two. The last
observation of this GRB was taken with the VLT $\sim$67 days after the
trigger, which would have been too soon for detecting the second
image, assuming $\sigma_{\nu}<250$ km/s. No late X-ray observations
were taken. In order to test this being a lensing system, direct
measurements of the velocity dispersion or galaxy mass are needed.

\subsection{Possible Candidates of Multiply Imaged GRBs}
\subsubsection{\emph{GRB021004}}
\begin{figure}[htbp!]
\begin{center}
\includegraphics[width=\columnwidth]{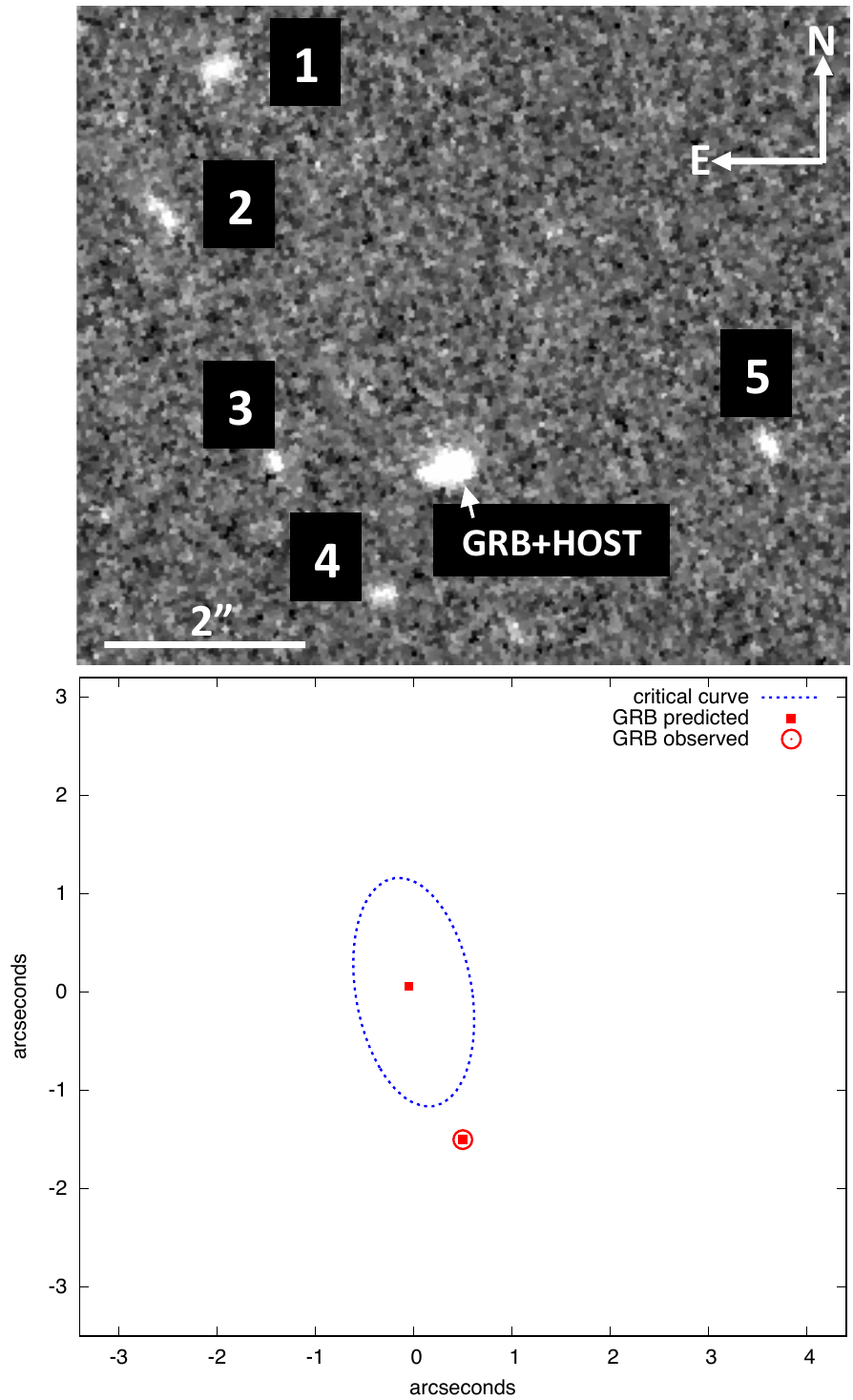}
\caption[short]{\emph{Top:} HST ACS/F606W field of GRB021004. The
  host complex and GRB are marked, and object 1-5 are the neighbouring
  galaxies. \emph{Bottom:} Possible GRAVLENS solution for
  GRB021004. The galaxy group was modeled as a SIE. The observed GRB
  is $\sim$25 times brighter than the second image, and arrives $\sim$550
  days earlier.  }
\label{fig:grb021004}
\end{center}
\end{figure}

While there is no exceptionally bright galaxy near this GRB, there are
five faint galaxies within $4.5^{\prime\prime}$ which could be a
part of a group (Fig.\ \ref{fig:grb021004}). At a redshift of 2.33,
this GRB had 2 strong absorbing systems at redshifts 1.38 and
1.60. 

\cite{0004-637X-633-1-317} studied the colors of the neighboring
galaxies and found the host galaxy to be different from its
surrounding. As gravitational lensing does not alter the color of
objects, this suggests that none of the other galaxies is a second
image of the host. Therefore, if modeling the group as an SIE and for the
case of a double image, the detected GRB must have been the first,
brighter one, and the second image of the host galaxy must not be
strongly magnified. Using the \emph{GRAVLENS} software we construct a
possible model of the group, having a common dark matter envelope
which can be approximated as a SIE. The model predicts multiple
images, with the second being too faint to detect the host. The center
of mass of the group model appears towards the more dense part of the
group, as expected. Since the time delay between the images is
$\sim$550 days, we would not expect to observe 2 images of the GRB at
the same time. Unfortunately, the HST observations stopped 53 days
after the trigger, which is less than the expected delay time for this
model. The host galaxy observations taken a year later would have
still been to early. Although this GRB had a bright optical afterglow,
with a relative demagnification of a factor of $\sim$25, the window of possible
detection is fairly narrow (several days).

Redshift confirmation for the galaxies is required to verify this is
indeed a bona fide group. In such a case, deeper images might reveal
other background sources and possibly multiple images of the
host. Phot-z analysis using the HST images finds galaxies 1-5 to be consistent with the
absorber's redshift within errors.

\subsubsection{\emph{GRB010222}}
\begin{figure}[htb]
\begin{center}
\includegraphics[width=\columnwidth]{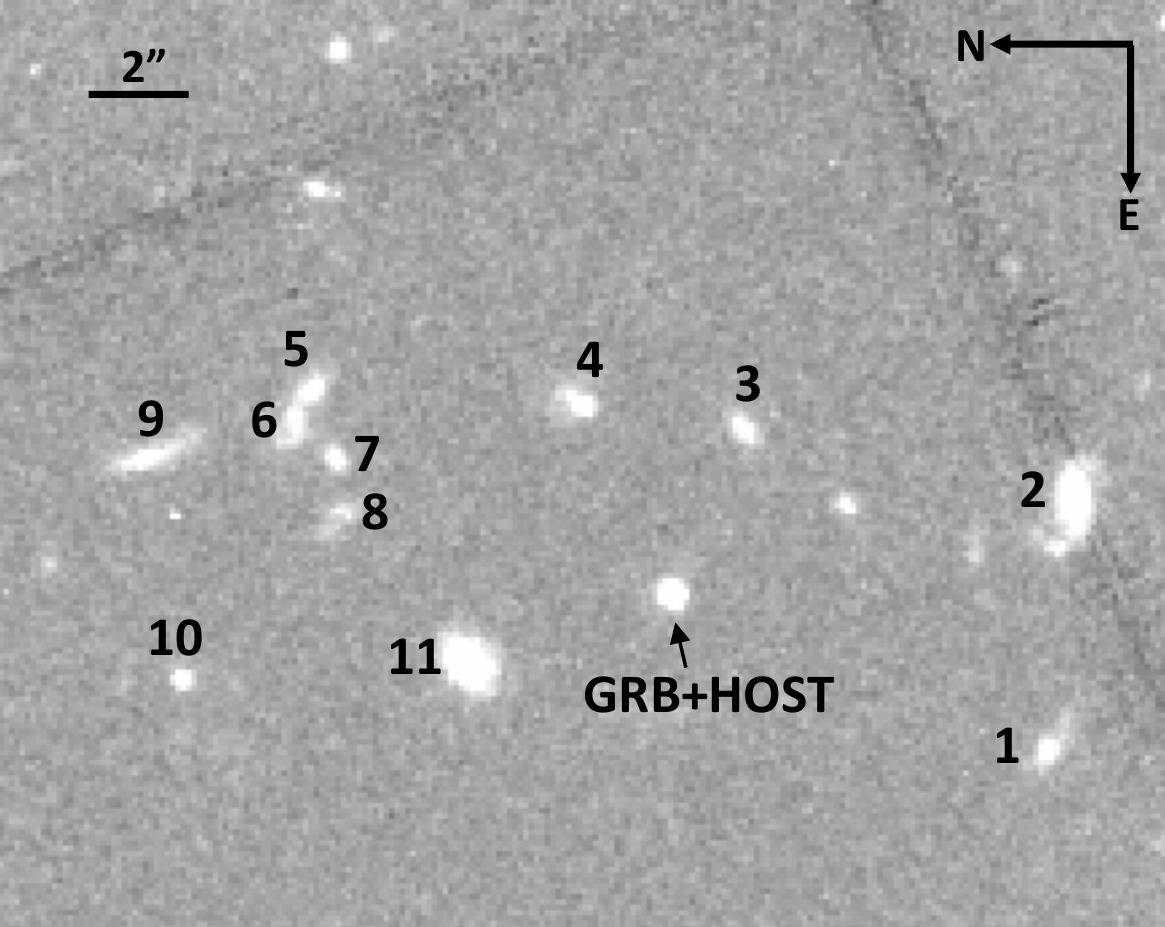}
\caption[short]{HST/WFPC2 F606W image of GRB010222. The host galaxy of the GRB
  is clearly visible. At least 11 galaxies are seen in the
  image, and could possibly be a part of a group (objects 1-11).}
\label{fig:grb010222}
\end{center}
\end{figure}

A strong MgII absorbing system was found at a redshift of 0.927 for
this z=1.477 GRB. While there was no single bright, close galaxy,
HST images show a crowded region with at least 11 clearly identified
galaxies which could, again, be a part of a group
(Fig. \ref{fig:grb010222}). To produce a second image, the velocity
dispersion of the group would need to be 250 km/s for a group
centered 0.5$^{\prime\prime}$ from the GRB, 350 km/s for
1$^{\prime\prime}$ between the GRB and group center, or 500 km/s for 2$^{\prime\prime}$.

Here again, one would need to confirm the galaxies are at the absorber
redshift as the first step to concluding that this group lensed the
GRB. As the SNR of these galaxies is low and the clumpiness of some
making it difficult to measure accurate photometry, we performs a
phot-z analysis on galaxies 1-4,9-11. The results, whilst carrying
large errors are consistent with the galaxies being at the absorber's redshift.

\subsubsection{\emph{GRB060206}}
\begin{figure}[htb]
\begin{center}
\includegraphics[width=\columnwidth]{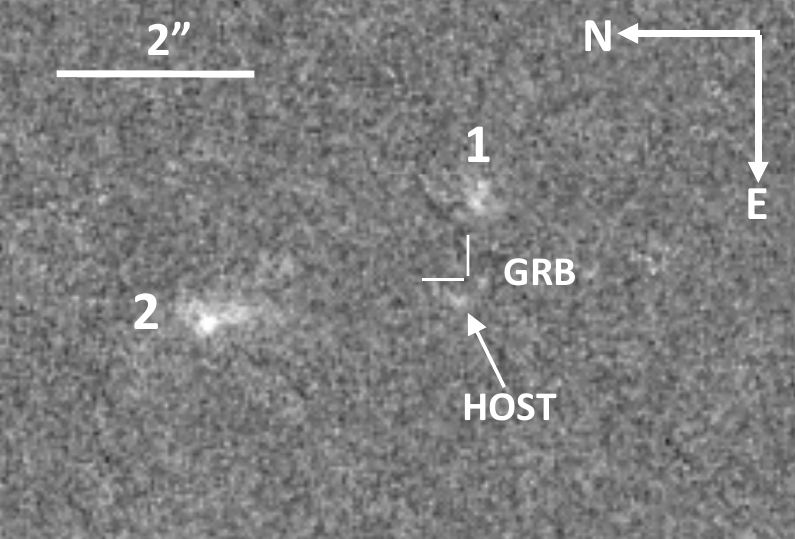}
\caption[short]{HST ACS/F814W field of GRB060206. Object 1 (2) is
  $\sim$ $0.96^{\prime\prime}$ ($2.45^{\prime\prime}$) from the host
  galaxy.}
\label{fig:grb060206}
\end{center}
\end{figure}

At a redshift of 4.048, GRB060206 exhibits 2 strong absorbing systems
at redshifts 2.26 and 1.48. An HST image reveals 2 galaxies near the
observed host (see Fig. \ref{fig:grb060206}), at distances
of $\sim$0.96$^{\prime\prime}$ for galaxy 1 and $\sim
2.45^{\prime\prime}$ for galaxy 2. The probability of finding an
alignment of two such galaxies along a random line of sight is 3\%, or
a 29\% chance of being observed in the 11 GRBs. The field was also
observed with GMOS r' on Gemini North (GN-2007A-Q-88), and galaxy 2
was detected with a magnitude 23.9$\pm$0.1
\citep{2008AA.489.37T}. Using the color information for galaxy 2, we
fit an SED model and estimate it could be a small starburst galaxy at
either absorber redshift, with $M_B\sim$ -16.35 (-17.46) for z=1.48
(2.26). Galaxy 2 would have to have a velocity dispersion larger than
325 km/s to create 2 images of this GRB. Since we do not expect such
velocities from a small starburst galaxy, we do not consider the
galaxy further.

With galaxy 1 only observed with one filter, it is impossible to determine
the galaxy type. Galaxy 1 would require an Einstein radius of $\sim
0.5^{\prime\prime}$ to lens the GRB, which is equivalent to $\sigma
\sim$ 200 (275) km/s at redshift 1.48 (2.26). A second image of the
GRB host would not necessarily be observed due to the small separation
from the lensing galaxy and the usual de-magnification of the second
image relative to the first image. Further photometry is required for
constraining the model of the galaxy and determining its lensing
feasibility.

\subsection{Unlikely Candidates of Multiply Imaged GRBs}
\subsubsection{\emph{GRB060418}}
\begin{figure}[htb]
\begin{center}
  \includegraphics[width=\columnwidth]{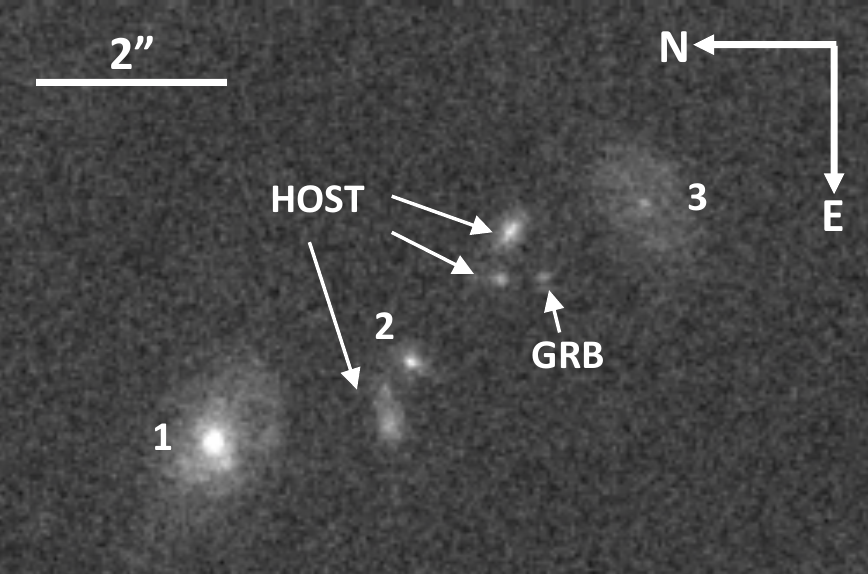}
  \caption[short]{HST/WFPC2 F775W image of GRB060418. The GRB and host
    locations are marked, 1-3 are nearby galaxies. Galaxy 1 was
    confirmed at the absorber z=0.656 redshift.}
\label{fig:grb060418}
\end{center}
\end{figure}

This GRB at z=1.49, with 3 strong absorbing systems at z=0.603, 0.656,
and 1.107, was well studied by \cite{0004-637X-701-2-1605}, who
identified a complicated host and 3 adjacent galaxies (figure
\ref{fig:grb060418}). In their paper, they identify galaxy 1 as that
responsible for the absorbing system at z=0.656, as its redshift was
confirmed with observed emission lines. Object 2 was assumed to be the
z=1.107 absorber and object 3 the z=0.603 absorber due to their
angular sizes. However, none of these later identifications were
confirmed via spectroscopy. Recently, \cite{2011arXiv1110.0487C}
studied the nearby galaxies and was able to spectroscoply confirm the
redshift of what is now identified as the host complex in figure
\ref{fig:grb060418}.

Due to the relatively large distance between the GRB and the probable
absorbers, we find lensing to be unlikely.

\subsubsection{\emph{GRB050820A}}
In their paper, \cite{0004-637X-691-1-152} study this GRB at z=2.615
and its strong absorbing systems at z=0.6915 and 1.4288. They identify 2
compact objects at separations of 1.3$^{\prime\prime}$ and
0.4$^{\prime\prime}$ from the GRB. In spectra taken lately by
\cite{2011arXiv1110.0487C}, the objects were found to be a part of the
GRB host. Therefore, we conclude that this GRB was not strongly lensed.

\subsubsection{\emph{GRB080319B}}
This GRB at a redshift of 0.9378, with one strong MgII intervening
system at z=0.7154 and 3 other weak systems, has 3 nearby
galaxies. Finding 3 galaxies within $4^{\prime\prime}$ is unlikely,
with the mean number of galaxies predicted to be 1.2, with a standard
deviation of 0.45. However, the faintness of the galaxies makes it
impossible to find a model which allows lensing. Although these data
were taken with only 2 filters, we were able to fit an SED model which
found them to be faint early type galaxies, with $-15.5<M_B<-12.5$
(assuming z=0.7154). Therefore, even if the galaxies are in the
absorber's redshift and not background galaxies, they could not be
massive enough to cause strong lensing. 

\subsubsection{\emph{GRB991216}}

\begin{figure}[htb]
\begin{center}
\includegraphics[width=0.7\columnwidth]{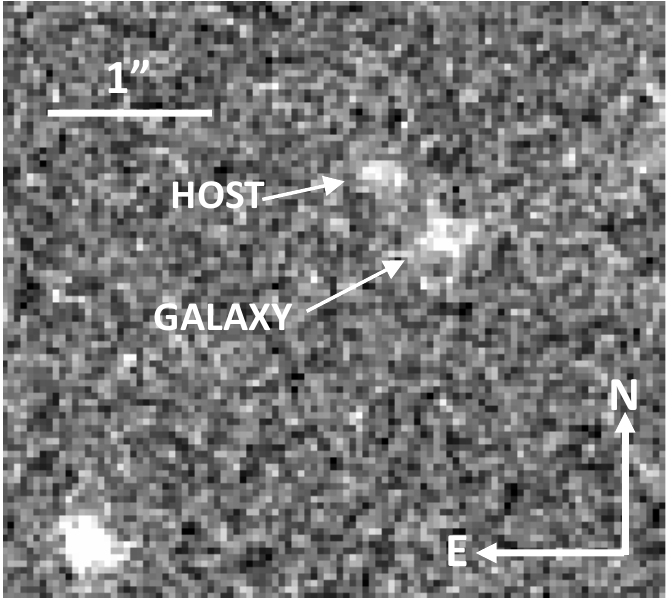}
\caption[short]{HST/STIS Long Pass image of GRB991216. The GRB host
  and the nearby irregular galaxy are marked.}
\label{fig:grb991216}
\end{center}
\end{figure}

A faint (R$\sim$24.5) irregular
galaxy\footnote{http://www.stsci.edu/$\sim$fruchter/GRB/991216/index.html}
lies $0.4^{\prime\prime}$ to the South of this GRB.  The GRB redshift
was 1.022 and the intervening systems were at 0.77 and 0.803. This
configuration requires that nearby galaxy to have a velocity
dispersion of at least 190 km/s, which is unlikely for a relatively
faint irregular galaxy. However, the proximity implies that even a
velocity dispersion of 70km/s would imply 15\%
magnification. Therefore, while not doubly lensed, this GRB (and its
host) are likely to be magnified if the nearby galaxy is at the
absorber's redshift.

\subsubsection{\emph{GRB020813}}
At a redshift of 1.255 this GRB shows a strong MgII absorber at a
redshift of 1.224. The ratio of $\frac{D_{ls}}{D_S}$ is $\sim$58 which
suggests that strong lensing is highly unlikely. The nearby bright
galaxy at a distance of 2.4$^{\prime\prime}$ away would require
$\sigma_{\nu}>1550$ km/s, which is unphysical.

\subsubsection{\emph{GRB050908}}
This GRB at z=3.55 with a strong MgII absorbing system at z= 1.548,
reveals a faint (F775W$\sim$26.6 mag) galaxy $\sim1^{\prime\prime}$
away. This alignment would require the galaxy to have a velocity
dispersion larger than 230 km/s. HST observations with different
filters are required to model the SED of this galaxy and conclude if
this high velocity dispersion is feasible.

\section{Discussion and Conclusions}
\label{discussion}

There is a well known excess of strong MgII absorbers towards GRBs
compared to quasar lines of sight. The most viable explanations for
this discrepancy are gravitational lensing of the GRB population, and
dust extinction of the population of quasars with strong absorbers. In
this paper we have identified a new difference between GRBs and random
lines of sight. We find that GRBs with strong MgII absorbers are found
closer to bright galaxies when compared with random lines of sight
that are restricted to the 60\% of sky nearest to foreground galaxies
(corresponding to the fraction of GRBs with strong MgII
absorption). This new property of the GRB population cannot be
explained by quasar extinction. On the other hand such a concentration
of GRBs around foreground galaxies is a natural consequence of
gravitational lensing. We therefore argue that strong gravitational
lensing is likely to be playing a role in explaining the discrepancy
between MgII absorbers towards GRBs and
quasars. \citeauthor{0004-637X-659-1-218}\ also found that GRBs with
MgII absorbing systems were slightly brighter (a factor of 1.7) than those
without, implying again that gravitational lensing could explain the
phenomenon, provided the GRB luminosity function was sufficiently
steep.

W11 predicted that 10\%-60\% of the 26 GRBs in the V09 sample should
have been multiply imaged if lensing is the explanation for the
enhanced MgII absorption towards GRBs relative to quasars. Assuming
that the GRBs with strong MgII systems (EW$>1$\AA) are the only ones
which are multiply-imaged, we would expect 1.9 - 11 of the 11 GRBs
studied to be strongly gravitationally lensed. If none of our sample
were plausibly gravitationally lensed, then this would effectively
rule out the gravitational lensing hypothesis. 

Quasars are observed to show strong lensing $\sim$0.1\% of the time,
and we expect, if the GRB and quasar samples are similarly
distributed, no GRBs in our sample to be lensed. It is important to
note that the W11 model predicts a lensing rate of less than 1\% for
both GRBs selected only in gamma-rays and for quasars. Thus, even the
null detection of multiple imaging in the 2700 GRBs observed by the
BATSE instrument, which could see a much larger fraction of the sky
than Swift at any given time, is not surprising (see
\citealt{2001ApJ...548..522P}).  

Our analysis reveals a high frequency of galaxies near GRB
sight-lines, with 2 suspected two-image systems (GRB020405 and
GRB030429), 2 possible galaxies groups that would result in multiple
images (GRB021004 and GRB010222), and several other less probable
candidates. Since only the two most likely candidates have spectroscopic
confirmation that the nearby galaxy lies at the absorber's redshift,
more data are needed to confirm the strong gravitational lensing
hypothesis. ñThis study, while not conclusively demonstrating lensing,
is consistent with what would be expected from the model of W11. If,
as described by Wyithe et al., the cumulative luminosity function of
the optical afterglow follows a power law of the form $\Psi_A(L_A)
\propto (L_A)^{-\alpha_A}$, where $L_A$ is the afterglow luminosity,
then our study indicates a value of $\alpha_A \sim 3.5$ or steeper.

Recently, \cite{2011arXiv1106.0692B} studied the correlation between
quasars, MgII absorbers and extinction, and found strong dependence of
E(B$-$V) on the absorber's equivalent width. Modeling the observed
difference between GRBs and quasars, they find dust obscuration to be a
significant factor in the MgII discrepancy at high equivalent
width. However, they acknowledge that it is unlikely to be the only
effect, and that the full explanation involves more than one
process. Our analysis compares the galaxy over-density towards GRBs
relative to a random line of sight and so is independent from
comparisons to other objects (e.g., quasars or blazars; see
\citealt{2011AA.525A.51B}). Therefore, both lensing and dust could be
of significance for solving the MgII problem.

Searching for galaxies near the lines of sight toward quasars with
strong MgII absorption would require detailed modeling and subtraction
of the quasar point spread function, which is beyond the scope of this
paper. The virtue of the GRB follow-up is the fading of the point
source, which affords a straight-forward assessment of the surrounding
sky. If follow-up studies of the most probable lenses do not
unambiguously rule in or out lensing, then a future direction could
include applying a similar analysis to a quasars sample. However, in
order to best test the gravitational lensing model, what is needed is
a real time study. Once an appropriate candidate is identified, deep
high-resolution images should follow, to search for the possible
lensing system. If one is found, lensing models should be applied to
predict if a second image is expected. If so, the location and time
delay of the next image can be measured. Depending on the error
expected in the time delay, which is proportional to $\sigma_{\nu}^3$,
an appropriate observational cadence can be put into action. Since the
second image is the inner one for the simple SIS/SIE models, it is
likely to suffer from stronger extinction by the lensing
galaxy. Moreover, the rapid fading in optical and the smaller
magnification of the second image compared to the first one might
require frequent observations in order to catch the second image while
it is observable. Therefore, we suggest conducting the search in
X-rays, which are less affected by passing through a galaxy.

The preferential lensing of GRBs would herald a shift in the study of
both gravitational lenses and GRBs. Prior knowledge of where and when
a GRB image will appear should allow coordinated observations from
earth and space to be scheduled, leading to unprecedented
multi-wavelength data. Such information would be of great assistance
in understanding progenitor and outflow properties. Moreover, accurate
measurements of the time delays between lensed images would constrain
cosmological models and inform studies of the dark matter distribution
associated with the lensing galaxy.

We hope that the conclusions drawn in this work will stimulate the
search for lensed GRBs (perhaps identified by their strong MgII
absorbers), and the associated tertiary images.

\section*{Acknowledgements}
We thank Jeremy Mould, John Kormendy and Stuart Sim for helpful conversations and
the anonymous referee for constructive comments. BPS acknowledges
financial support through ARC Laureate Fellowship Grant FL0992131.

Based on observations made with the European Southern Observatory
telescopes obtained from the ESO/ST-ECF Science Archive Facility.

Some of the data presented in this paper were obtained from the
Multimission Archive at the Space Telescope Science Institute
(MAST). STScI is operated by the Association of Universities for
Research in Astronomy, Inc., under NASA contract NAS5-26555. Support
for MAST for non-HST data is provided by the NASA Office of Space
Science via grant NNX09AF08G and by other grants and contracts.

Based on observations obtained at the Gemini Observatory (acquired
through the Gemini Science Archive), which is operated by the
Association of Universities for Research in Astronomy, Inc., under a
cooperative agreement with the NSF on behalf of the Gemini
partnership: the National Science Foundation (United States), the
Science and Technology Facilities Council (United Kingdom), the
National Research Council (Canada), CONICYT (Chile), the Australian
Research Council (Australia), Ministério da Ciência e Tecnologia
(Brazil) and Ministerio de Ciencia, Tecnología e Innovación Productiva
(Argentina).

 \bibliographystyle{hapj}

 \bibliography{ms}

\begin{thebibliography}{38}
\expandafter\ifx\csname natexlab\endcsname\relax\def\natexlab#1{#1}\fi

\bibitem[{{Beckwith} {et~al.}(2006){Beckwith}, {Stiavelli}, {Koekemoer},
  {Caldwell}, {Ferguson}, {Hook}, {Lucas}, {Bergeron}, {Corbin}, {Jogee},
  {Panagia}, {Robberto}, {Royle}, {Somerville}, \& {Sosey}}]{2006AJ.132.1729B}
{Beckwith}, S.~V.~W. {et~al.} 2006, \aj, 132, 1729

\bibitem[{{Bergeron} {et~al.}(2011){Bergeron}, {Boiss{\'e}}, \&
  {M{\'e}nard}}]{2011AA.525A.51B}
{Bergeron}, J., {Boiss{\'e}}, P., \& {M{\'e}nard}, B. 2011, \aap, 525, A51+

\bibitem[{{Bertin} \& {Arnouts}(1996)}]{1996AAS.117.393B}
{Bertin}, E., \& {Arnouts}, S. 1996, \aap, 117, 393

\bibitem[{{Binney} \& {Tremaine}(1987)}]{binney1988galactic}
{Binney}, J., \& {Tremaine}, S. 1987, {Galactic dynamics} (Princeton University
  Press), 230

\bibitem[{{Bornancini} {et~al.}(2004){Bornancini}, {Mart{\'{\i}}nez}, {Lambas},
  {Le Floc'h}, {Mirabel}, \& {Minniti}}]{2004ApJ...614...84B}
{Bornancini}, C.~G., {Mart{\'{\i}}nez}, H.~J., {Lambas}, D.~G., {Le Floc'h},
  E., {Mirabel}, I.~F., \& {Minniti}, D. 2004, \apj, 614, 84

\bibitem[{{Brammer} {et~al.}(2008){Brammer}, {van Dokkum}, \&
  {Coppi}}]{2008ApJ.686.1503B}
{Brammer}, G.~B., {van Dokkum}, P.~G., \& {Coppi}, P. 2008, \apj, 686, 1503,
  0807.1533

\bibitem[{{Budzynski} \& {Hewett}(2011)}]{2011arXiv1106.0692B}
{Budzynski}, J.~M., \& {Hewett}, P.~C. 2011, ArXiv, 1106.0692

\bibitem[{{Calzetti} {et~al.}(1994){Calzetti}, {Kinney}, \&
  {Storchi-Bergmann}}]{CKSB94}
{Calzetti}, D., {Kinney}, A.~L., \& {Storchi-Bergmann}, T. 1994, \apj, 429, 582

\bibitem[{{Calzetti} {et~al.}(1996){Calzetti}, {Kinney}, \&
  {Storchi-Bergmann}}]{Calzetti96}
------. 1996, \apj, 458, 132

\bibitem[{{Campisi} {et~al.}(2009){Campisi}, {De Lucia}, {Li}, {Mao}, \&
  {Kang}}]{2009MNRAS.400.1613C}
{Campisi}, M.~A., {De Lucia}, G., {Li}, L.-X., {Mao}, S., \& {Kang}, X. 2009,
  \mnras, 400, 1613

\bibitem[{{Chen}(2011)}]{2011arXiv1110.0487C}
{Chen}, H.-W. 2011, ArXiv, 1110.0487

\bibitem[{{Chen} {et~al.}(2010){Chen}, {Helsby}, {Gauthier}, {Shectman},
  {Thompson}, \& {Tinker}}]{2010ApJ.714.1521C}
{Chen}, H.-W., {Helsby}, J.~E., {Gauthier}, J.-R., {Shectman}, S.~A.,
  {Thompson}, I.~B., \& {Tinker}, J.~L. 2010, \apj, 714, 1521

\bibitem[{Chen {et~al.}(2009)Chen, Perley, Pollack, Prochaska, Bloom,
  Dessauges-Zavadsky, Pettini, Lopez, Dall'aglio, \&
  Becker}]{0004-637X-691-1-152}
Chen, H.-W. {et~al.} 2009, \apj, 691, 152

\bibitem[{{Churchill} {et~al.}(2005){Churchill}, {Kacprzak}, \&
  {Steidel}}]{2005pgqa.conf.24C}
{Churchill}, C.~W., {Kacprzak}, G.~G., \& {Steidel}, C.~C. 2005, in IAU Colloq.
  199: Probing Galaxies through Quasar Absorption Lines, ed. {P.~Williams,
  C.-G.~Shu, \& B.~Menard}, 24--41

\bibitem[{{Coe} {et~al.}(2006){Coe}, {Ben{\'{\i}}tez}, {S{\'a}nchez}, {Jee},
  {Bouwens}, \& {Ford}}]{2006AJ.132.926C}
{Coe}, D., {Ben{\'{\i}}tez}, N., {S{\'a}nchez}, S.~F., {Jee}, M., {Bouwens},
  R., \& {Ford}, H. 2006, \aj, 132, 926

\bibitem[{{Cucchiara} {et~al.}(2009){Cucchiara}, {Jones}, {Charlton}, {Fox},
  {Einsig}, \& {Narayanan}}]{2009ApJ...697..345C}
{Cucchiara}, A., {Jones}, T., {Charlton}, J.~C., {Fox}, D.~B., {Einsig}, D., \&
  {Narayanan}, A. 2009, \apj, 697, 345

\bibitem[{Fynbo {et~al.}(2005)Fynbo, Gorosabel, Smette, Fruchter, Hjorth,
  Pedersen, Levan, Burud, Sahu, Vreeswijk, Bergeron, Kouveliotou, Tanvir,
  Thorsett, Wijers, Cerón, Castro-Tirado, Garnavich, Holland, Jakobsson,
  Møller, Nugent, Pian, Rhoads, Thomsen, Watson, \&
  Woosley}]{0004-637X-633-1-317}
Fynbo, J. P.~U. {et~al.} 2005, \apj, 633, 317

\bibitem[{{Hook} {et~al.}(2004){Hook}, {J{\o}rgensen}, {Allington-Smith},
  {Davies}, {Metcalfe}, {Murowinski}, \& {Crampton}}]{2004PASP.116.425H}
{Hook}, I.~M., {J{\o}rgensen}, I., {Allington-Smith}, J.~R., {Davies}, R.~L.,
  {Metcalfe}, N., {Murowinski}, R.~G., \& {Crampton}, D. 2004, \pasp, 116, 425

\bibitem[{{Jakobsson} {et~al.}(2004){Jakobsson}, {Hjorth}, {Fynbo},
  {Weidinger}, {Gorosabel}, {Ledoux}, {Watson}, {Bj{\"o}rnsson}, {Gudmundsson},
  {Wijers}, {M{\"o}ller}, {Pedersen}, {Sollerman}, {Henden}, {Jensen},
  {Gilmore}, {Kilmartin}, {Levan}, {Castro Cer{\'o}n}, {Castro-Tirado},
  {Fruchter}, {Kouveliotou}, {Masetti}, \& {Tanvir}}]{2004AA.427.785J}
{Jakobsson}, P. {et~al.} 2004, \aap, 427, 785

\bibitem[{Kacprzak {et~al.}(2007)Kacprzak, Churchill, Steidel, Murphy, \&
  Evans}]{0004-637X-662-2-909}
Kacprzak, G.~G., Churchill, C.~W., Steidel, C.~C., Murphy, M.~T., \& Evans,
  J.~L. 2007, \apj, 662, 909

\bibitem[{{Keeton}(2001)}]{2001astro.ph..2340K}
{Keeton}, C.~R. 2001, ArXiv, 0102340

\bibitem[{{Kinney} {et~al.}(1996){Kinney}, {Calzetti}, {Bohlin}, {McQuade},
  {Storchi-Bergmann}, \& {Schmitt}}]{Kinney96}
{Kinney}, A.~L., {Calzetti}, D., {Bohlin}, R.~C., {McQuade}, K.,
  {Storchi-Bergmann}, T., \& {Schmitt}, H.~R. 1996, \apj, 467, 38

\bibitem[{{Mannucci} {et~al.}(2001){Mannucci}, {Basile}, {Poggianti},
  {Cimatti}, {Daddi}, {Pozzetti}, \& {Vanzi}}]{IRTemplates}
{Mannucci}, F., {Basile}, F., {Poggianti}, B.~M., {Cimatti}, A., {Daddi}, E.,
  {Pozzetti}, L., \& {Vanzi}, L. 2001, \mnras, 326, 745

\bibitem[{{Masetti} {et~al.}(2003){Masetti}, {E. Palazzi}, {E. Pian}, {A.
  Simoncelli}, {L. K. Hunt}, {E. Maiorano}, {A. Levan}, {L. Christensen}, {E.
  Rol}, {S. Savaglio}, {R. Falomo}, {A. J. Castro-Tirado}, {J. Hjorth}, {A.
  Delsanti}, {M. Pannella}, {V. Mohan}, {S. B. Pandey}, {R. Sagar}, {L. Amati},
  {I. Burud}, {J. M. Castro Cer\'on}, {F. Frontera}, {A. S. Fruchter}, {J. P.
  U. Fynbo}, {J. Gorosabel}, {L. Kaper}, {S. Klose}, {C. Kouveliotou}, {L.
  Nicastro}, {H. Pedersen}, {J. Rhoads}, {I. Salamanca}, {N. Tanvir}, {P. M.
  Vreeswijk}, {R. A. M. J. Wijers}, \& {E. P. J. van den Heuvel}}]{grb020405}
{Masetti} {et~al.} 2003, A\&A, 404, 465

\bibitem[{{Miller} {et~al.}(2011){Miller}, {Bundy}, {Sullivan}, {Ellis}, \&
  {Treu}}]{2011arXiv1102.3911M}
{Miller}, S.~H., {Bundy}, K., {Sullivan}, M., {Ellis}, R.~S., \& {Treu}, T.
  2011, ArXiv, 1102.3911

\bibitem[{{Moster} {et~al.}(2011){Moster}, {Somerville}, {Newman}, \&
  {Rix}}]{cosmic}
{Moster}, B.~P., {Somerville}, R.~S., {Newman}, J.~A., \& {Rix}, H.-W. 2011,
  \apj, 731, 113

\bibitem[{{Nestor} {et~al.}(2005){Nestor}, {Turnshek}, \& {Rao}}]{Nestor2005}
{Nestor}, D.~B., {Turnshek}, D.~A., \& {Rao}, S.~M. 2005, \apj, 628, 637

\bibitem[{Pollack {et~al.}(2009)Pollack, Chen, Prochaska, \&
  Bloom}]{0004-637X-701-2-1605}
Pollack, L.~K., Chen, H.-W., Prochaska, J.~X., \& Bloom, J.~S. 2009, \apj, 701,
  1605

\bibitem[{{Porciani} \& {Madau}(2001)}]{2001ApJ...548..522P}
{Porciani}, C., \& {Madau}, P. 2001, \apj, 548, 522

\bibitem[{Porciani {et~al.}(2007)Porciani, Viel, \&
  Lilly}]{0004-637X-659-1-218}
Porciani, C., Viel, M., \& Lilly, S.~J. 2007, \apj, 659, 218

\bibitem[{Prochter {et~al.}(2006)Prochter, Prochaska, Chen, Bloom,
  Dessauges-Zavadsky, Foley, Lopez, Pettini, Dupree, \&
  Guhathakurta}]{1538-4357-648-2-L93}
Prochter, G.~E. {et~al.} 2006, \apjl, 648, L93

\bibitem[{{Schneider} {et~al.}(1999){Schneider}, {Ehlers}, \&
  {Falco}}]{Schneider}
{Schneider}, P., {Ehlers}, J., \& {Falco}, E.~E. 1999, Gravitational Lenses
  (Berlin: Springer)

\bibitem[{{Tejos} {et~al.}(2009){Tejos}, {Lopez}, {Prochaska}, {Bloom}, {Chen},
  {Dessauges-Zavadsky}, \& {Maureira}}]{Tejos2009}
{Tejos}, N., {Lopez}, S., {Prochaska}, J.~X., {Bloom}, J.~S., {Chen}, H.-W.,
  {Dessauges-Zavadsky}, M., \& {Maureira}, M.~J. 2009, \apj, 706, 1309

\bibitem[{{Tejos} {et~al.}(2007){Tejos}, {Lopez}, {Prochaska}, {Chen}, \&
  {Dessauges-Zavadsky}}]{Tejos2007}
{Tejos}, N., {Lopez}, S., {Prochaska}, J.~X., {Chen}, H.-W., \&
  {Dessauges-Zavadsky}, M. 2007, \apj, 671, 622

\bibitem[{{Thone} {et~al.}(2008){Thone}, {Wiersema}, {Ledoux}, {Starling}, {de
  Ugarte Postigo}, {Levan}, {Fynbo}, {Curran}, {Gorosabel}, {van der Horst},
  {Llorente}, {Rol}, {Tanvir}, {Vreeswijk}, {Wijers}, \&
  {Kewley}}]{2008AA.489.37T}
{Thone}, C.~C. {et~al.} 2008, \aap, 489, 37

\bibitem[{{Vergani} {et~al.}(2009){Vergani}, {Petitjean}, {Ledoux},
  {Vreeswijk}, {Smette}, \& {Meurs}}]{2009AA.503.771V}
{Vergani}, S.~D., {Petitjean}, P., {Ledoux}, C., {Vreeswijk}, P., {Smette}, A.,
  \& {Meurs}, E.~J.~A. 2009, Astronomy \& Astrophysics, 503, 771

\bibitem[{Wyithe {et~al.}(2011)Wyithe, Oh, \& Pindor}]{2010arXiv1004.2081W}
Wyithe, J. S.~B., Oh, S.~P., \& Pindor, B. 2011, \mnras, 414, 209

\bibitem[{{Wyithe} {et~al.}(2003){Wyithe}, {Winn}, \&
  {Rusin}}]{2003ApJ...583...58W}
{Wyithe}, J.~S.~B., {Winn}, J.~N., \& {Rusin}, D. 2003, \apj, 583, 58

\end{thebibliography}
\end{document}